\documentclass[aps,prb,twocolumn, superscriptaddress, floatfix, longbibliography]{revtex4-1}
\usepackage{graphicx}
\usepackage{dcolumn}
\usepackage{bm}
\usepackage{commath}
\usepackage{turnstile}
\usepackage{amsmath,amssymb}
\usepackage{xcolor}
\usepackage{url}
\usepackage{cases}
\usepackage{epsfig}
\usepackage{epstopdf}
\usepackage{float}
\usepackage[hidelinks]{hyperref}
\newcommand{\cor}[1]{\textcolor{black}{#1}}
\begin{document}

\title{Resonant x-ray scattering study of charge-density wave correlations in YBa$_{2}$Cu$_{3}$O$_{6+x}$ under uniaxial stress}
\author{S. Nakata}
\affiliation{Max Planck Institute for Solid State Research, Heisenbergstra\ss e 1, D-70569 Stuttgart, Germany}
\author{D. Betto}
\affiliation{Max Planck Institute for Solid State Research, Heisenbergstra\ss e 1, D-70569 Stuttgart, Germany}
\author{E. Schierle}
\affiliation{Helmholtz-Zentrum Berlin f\"{u}r Materialien und Energie, Wilhelm-Conrad-R\"{o}ntgen-Campus BESSY II, Albert-Einstein-Str. 15, 12489 Berlin, Germany}
\author{S. Hameed}
\affiliation{Max Planck Institute for Solid State Research, Heisenbergstra\ss e 1, D-70569 Stuttgart, Germany}
\author{Y. Liu}
\affiliation{Max Planck Institute for Solid State Research, Heisenbergstra\ss e 1, D-70569 Stuttgart, Germany}
\author{H.-H. Kim}
\affiliation{Max Planck Institute for Solid State Research, Heisenbergstra\ss e 1, D-70569 Stuttgart, Germany}
\author{S. M. Souliou}
\affiliation{Institute for Quantum Materials and Technologies, Karlsruhe Institute of Technology, Kaiserstr. 12, 76131 Karlsruhe, Germany}
\author{T. Lacmann}
\affiliation{Institute for Quantum Materials and Technologies, Karlsruhe Institute of Technology, Kaiserstr. 12, 76131 Karlsruhe, Germany}
\author{K. F\"{u}rsich}
\affiliation{Max Planck Institute for Solid State Research, Heisenbergstra\ss e 1, D-70569 Stuttgart, Germany}
\author{T. Loew}
\affiliation{Max Planck Institute for Solid State Research, Heisenbergstra\ss e 1, D-70569 Stuttgart, Germany}
\author{E. Weschke}
\affiliation{Helmholtz-Zentrum Berlin f\"{u}r Materialien und Energie, Wilhelm-Conrad-R\"{o}ntgen-Campus BESSY II, Albert-Einstein-Str. 15, 12489 Berlin, Germany}
\author{A. P. Mackenzie}
\affiliation{Max Planck Institute for Chemical Physics of Solids, N\"{o}thnitzer Stra\ss e 40, 01187 Dresden, Germany}
\affiliation{Scottish Universities Physics Alliance, School of Physics and Astronomy, University of
St Andrews, St Andrews KY16 9SS, UK}
\author{C. W. Hicks}
\affiliation{Max Planck Institute for Chemical Physics of Solids, N\"{o}thnitzer Stra\ss e 40, 01187 Dresden, Germany}
\author{M. Le Tacon}
\affiliation{Institute for Quantum Materials and Technologies, Karlsruhe Institute of Technology, Kaiserstr. 12, 76131 Karlsruhe, Germany}
\author{B. Keimer}
\affiliation{Max Planck Institute for Solid State Research, Heisenbergstra\ss e 1, D-70569 Stuttgart, Germany}
\author{M. Minola}
\affiliation{Max Planck Institute for Solid State Research, Heisenbergstra\ss e 1, D-70569 Stuttgart, Germany}

\begin{abstract}
We report a comprehensive study of the uniaxial stress response of charge-density-wave (CDW) correlations in detwinned single crystals of the high temperature superconductor YBa$_2$Cu$_3$O$_{6+x}$ (YBCO$_{6+x}$) with $0.40 \leq x \leq 0.93$ (hole-doping levels $0.072 \leq p \leq 0.168$) by means of Cu $L_3$-edge resonant energy-integrated x-ray scattering (REXS). We show that the influence of uniaxial stress is strongly doping dependent: the quasi-two-dimensional CDW is enhanced by in-plane uniaxial stress in a wide hole doping range ($0.45 \leq x \leq 0.80$), but only barely affected in the most underdoped and optimally doped samples ($x = 0.40$ and 0.93), where the CDW correlation length is minimal. A stress-induced three-dimensionally long-range ordered (3D) CDW was observed only in YBCO$_{6.50}$ and YBCO$_{6.67}$. The temperature dependence of the 3D CDW clearly indicates a strong competition with superconductivity. Based on the systematic strain-, doping-, and temperature-dependent REXS measurements reported here, we discuss the relationship between charge order and superconductivity in YBCO$_{6+x}$ and other cuprates.
\end{abstract}

\maketitle

\section{\label{sec:level1}Introduction}
The interplay among spin, charge, and lattice degrees of freedom of electrons in quantum materials shapes complex phase diagrams in which a number of competing ordering phenomena emerge \cite{RevModPhys.70.1039,Keimer2017The-physic}. Tuning the charge carrier concentration and applying external fields have uncovered a complex landscape of electronic phases in many quantum materials \cite{Basov2017Towards-pr}, including particularly the copper oxide high-temperature superconductors \cite{Keimer2015From-quantum-ma}. Removing electrons from the undoped Mott insulating CuO$_2$ planes supplants long-range ordered antiferromagnetism and gives rise to superconductivity. Charge-density-wave (CDW) correlations are observed in all cuprate families at moderate doping levels. Their intensity generally increases upon cooling in the normal state, and decreases below the superconducting transition temperature $T_\text{c}$, manifesting a competition between both ordering phenomena \cite{Frano_2020}. Many aspects of charge ordering are, however, different among cuprate families and influenced by doping-induced disorder. \cor{In particular, families with higher chemical disorder such a Bi-based cuprates tend to have an overall weaker charge order signal than that of cleaner systems including Y-based cuprates. This is reflected in less intense, broader peaks, indicating shorter correlation length, and typically as a more monotonic temperature dependence of the charge order signal, whose competition with superconductivity below $T_\text{c}$ also becomes less evident.}

External control parameters allow the modulation of the interplay between CDW order and superconductivity without introducing disorder. The application of external magnetic fields weakens superconductivity and enhances quasi-two-dimensional (quasi-2D) charge order below the zero-field $T_\text{c}$ \cite{Chang2012Direct-obs}. In underdoped YBa$_2$Cu$_3$O$_{6+x}$ (YBCO$_{6+x}$), magnetic fields above $\sim$15 T induce a long-range ordered CDW with three-dimensional (3D) correlations between CuO$_2$ bilayers in adjacent unit cells \cor{(with correlation lengths of order one unit cell along the c-axis)} \cite{Gerber949,Chang2016Magnetic-f}. In the presence of a magnetic field, the normal state transport coefficients near $T$ = 0 also exhibit quantum oscillations characteristic of small electron pockets \cite{Doiron-Leyraud2007Quantum-os}, and a sign reversal of the Hall coefficient \cite{LeBoeuf2007Electron-p} indicating a Fermi surface reconstruction triggered by charge ordering \cite{PhysRevB.100.045128}. While it is possible to apply fields of the order of 100 T in transport experiments, an analogous setup for spectroscopic studies is currently not available. Hydrostatic pressure is another continuously tunable parameter that modulates the interaction between charge order and superconductivity. However, the comparison between transport and x-ray scattering experiments under the same amount of pressure has not yet yielded a clear picture, partly reflecting difficulties of performing scattering experiments in the presence of the diamond anvil cell surrounding the sample \cite{PhysRevB.97.020503,PhysRevB.100.094502,PhysRevB.98.064513,PhysRevB.97.174508}. 

Recent technical developments allowing the application of highly homogeneous uniaxial stress have yielded additional insights into the competing phases of quantum materials \cite{Hicks283}. X-ray scattering measurements on YBCO$_{6.67}$ have shown that uniaxial stress strongly amplifies 2D charge order and induces 3D charge order\cor{(with correlation lengths of tens of unit cells along the c-axis)}, analogous to the magnetic field effects \cite{Kim1040,PhysRevLett.126.037002,Vinograd2024Using-stra}. Complementary transport measurements are also possible \cite{PhysRevLett.120.076602}, and allow direct comparison of microscopic correlation functions accessible via scattering probes with macroscopic transport coefficients under identical experimental conditions \cite{Nakata2022Normal-sta}. 

Despite the remarkable response of the CDW to uniaxial stress in YBCO$_{6+x}$, this effect has so far been studied mostly only at one particular doping level ($x$ = 0.67 corresponding to $p$ = 0.12 $\sim$ 1/8). Two other close doping levels were very recently measured only via nonresonant x-ray diffraction \cite{Vinograd2024Using-stra}, calling for a thorough investigation of the doping-dependence of this response and in particular of the stability range of the stress-induced 3D CDW  in the phase diagram. To address these issues, we performed Cu $L_3$ resonant energy-integrated x-ray scattering (REXS) measurements on detwinned single crystals of YBCO$_{6+x}$. To comprehensively investigate the stress response of charge ordering, we studied three doping levels ($x$ = 0.40, 0.45, and 0.50) lower than  $p$ = 0.12 and two doping levels ($x$ = 0.80 and 0.93) approaching optimal doping. We find that the 2D charge modulation perpendicular to the uniaxial stress axis is enhanced in a wide range of doping levels ($x = 0.45, 0.50, $ and 0.80). However the same type of amplification is not discernible in the very underdoped and optimally doped samples ($x = 0.40$ and 0.93). Moreover, we find that the 3D CDW along the $b$-axis is induced by uniaxial stress ($\sim$ 0.7 \%) in YBCO$_{6.50}$ in a similar fashion to YBCO$_{6.67}$, but this transition was not observed at any other doping level at the maximum stress currently achievable. Our findings thus indicate that the stress-induced 3D CDW correlations are strongly peaked in the vicinity of $p = 1/8$. Based on our systematic doping-, strain-, and temperature-dependent measurements of CDWs, \cor{we discuss how uniaxial stress enhances two-dimensional CDW across various doping levels, induces a three-dimensional CDW particularly in the vicinity of $p$ = 1/8, and reveals a competitive relationship between charge order and superconductivity symmetric around $p$ = 1/8 and with temperature dependencies highlighting the distinct responses of 2D and 3D CDWs to external perturbations.}

\section{\label{sec:level2}Experimental methods}
\begin{figure}
  \includegraphics[width=\columnwidth]{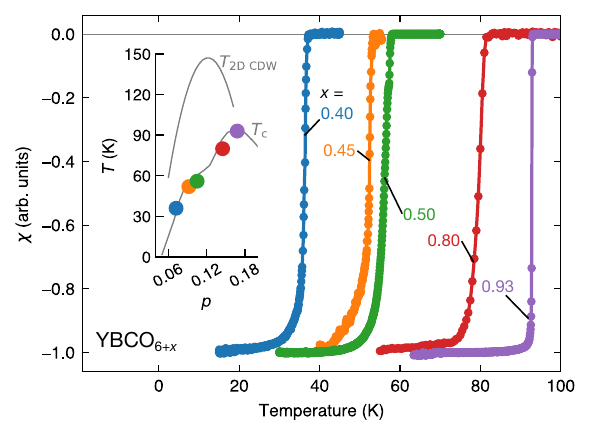}\\
  \caption{Magnetometry of YBCO$_{6+x}$ samples investigated in the present work. The diamagnetic response is normalized for clarity. Inset: the temperature-doping ($T$-$p$) phase diagram of YBCO$_{6+x}$. The onset of 2D CDW (gray curve) is based on the data reported in Ref. \onlinecite{PhysRevB.90.054513}. Dots indicate $T_\text{c}$ and $p$ of the samples shown in Table \ref{table:YBCO_REXS}.}
  \label{fig:Tc}
\end{figure}

YBCO$_{6+x}$ single crystals were synthesized using the self-flux method \cite{LIANG199251}. We studied five oxygen stoichiometries ($x$ = 0.40, 0.45, 0.50, 0.80, and 0.93), which were tuned by annealing crystals with well controlled oxygen partial pressure and temperature \cite{doi:10.1111/j.1151-2916.1989.tb05978.x}. The corresponding hole doping levels are shown in Table \ref{table:YBCO_REXS}. Subsequently, all crystals were detwinned by heating under uniaxial pressure \cite{LIN200427}. The $c$-axis lattice parameter at room temperature was determined with x-ray diffraction measurements and $T_{\rm c}$ was defined as the midpoint of the diamagnetic response in magnetic susceptibility measurements (Table 1 and Fig. \ref{fig:Tc}). To maximize the homogeneity of uniaxial stress applied during the REXS measurements, the single crystals were cut into a needle shape whose dimensions were typically $a \times b \times c = 1.5 \times 0.2 \times 0.1$ mm$^3$ as in the case of previous scattering experiments under stress \cite{Kim1040,PhysRevLett.126.037002,Kim2022Giant-stre}. Compressive stress was applied along the long axis of these needles. We pressurized the samples using the same piezoelectric-based apparatus used for previous x-ray scattering experiments on YBCO$_{6.67}$ at other synchrotron facilities \cite{Kim1040,PhysRevLett.126.037002}.

The REXS experiments were performed using linearly polarized photons produced by the elliptical undulator of the UE46-PGM1 beamline at the BESSY II synchrotron of the Helmholtz Zentrum Berlin. All spectra were collected at the Cu $L_3$ absorption edge (931.8 eV) with $\sigma$ polarization of the incident x-ray beam to maximize the charge contribution to the cross section in the net REXS signal \cite{Ghiringhelli821}. \cor{In the present setup, the wavevector resolution was set to be of the order of ~0.002 r.l.u. or better.} In principle, the REXS signal comprises contributions from inelastic scattering as well \footnote{\cor{This is in part why we use REXS as the abbreviation of Resonant Energy-integrated X-Ray Scattering whereas REXS is often used as the abbreviation of Resonant Elastic X-Ray Scattering in other literature.}}. In the case of YBCO$_{6+x}$, the dominant contribution to the inelastic signal is a $dd$ excitation, which could be sensitive to the symmetry of the lattice structure. However, the $dd$ excitations are known to show only a monotonic momentum dependence and no appreciable stress dependence, as observed in previous resonant inelastic x-ray scattering (RIXS) measurements on YBCO$_{6.67}$ \cite{PhysRevLett.126.037002}. We thus attribute the momentum and stress dependent REXS signal in the present study to static CDW modulations. Throughout this study, the scattering plane of the photons was parallel to the $bc$ plane and the uniaxial stress was applied along the $a$-axis (Fig. \ref{fig:2dcdw} (a)). 

The cross section of the incident x-ray beam (100 $\times$ 50 \textmu m$^2$) was smaller than the strained sample area (600 $\times$ 200 \textmu m$^2$), thus ensuring a homogeneous strain within the probed sample volume. The strain level was controlled by monitoring the capacitance of a parallel-plate capacitor incorporated in the stress device \cite{doi:10.1063/1.4881611} and calibrated against x-ray diffraction data collected on nearly identical needles from the same batch of crystals \cite{Vinograd2024Using-stra} (see the Appendix). Moreover, we monitored the shift of the (002) Bragg peak upon application of in-plane uniaxial stress in order to gauge the expansion of the out-of-plane lattice parameter due to Poisson's effect and to have a second {\it in-situ} diagnostic of strain, in addition to the capacitor. With this calibration, the experimental results reported here are fully compatible with those of Ref. \onlinecite{Vinograd2024Using-stra}, whereas the strain levels in Refs. \onlinecite{Kim1040,PhysRevLett.126.037002} were overestimated.

\begin{table}
  \begin{center}
    \begin{tabular}{ccccc}
    $x$ in YBCO$_{6+x}$ \qquad & $c$ (\AA) & $p$ & $T_\text{c}$ (K) & Structure\\\hline\hline
    0.40& 11.772 & 0.072 & 36  & O-II\\
    0.45& 11.752 & 0.092 &  52 & O-II\\
    0.50& 11.739 & 0.105 & 56 &  O-II\\
    0.67& 11.725 & 0.12 & 65  & O-VIII\\
    0.80& 11.708 & 0.145 & 80 &  O-III\\
    0.93 & 11.694 & 0.168 & 93 &  O-I\\\hline
    \end{tabular}
  \end{center}
  \caption{List of the YBCO$_{6+x}$ crystals investigated in the present study. The $c$-axis lattice parameter $c$ was determined by x-ray diffraction measurements at room temperature. The hole doping level $p$ was estimated based on $c$ using the empirical relationship shown in Ref. \onlinecite{PhysRevB.73.180505}. $T_\text{c}$ was determined by magnetometry as the midpoint of the diamagnetic response. The oxygen ordering pattern in the CuO-chain layers is also listed (see, for example, Ref. \onlinecite{PhysRevB.68.104515}). Previously reported data on YBCO$_{6.67}$  \cite{Kim1040,PhysRevLett.126.037002} are listed for comparison.}
  \label{table:YBCO_REXS}
\end{table}

\section{\label{sec:level3}Results}
\begin{figure}
  \includegraphics[width=\columnwidth]{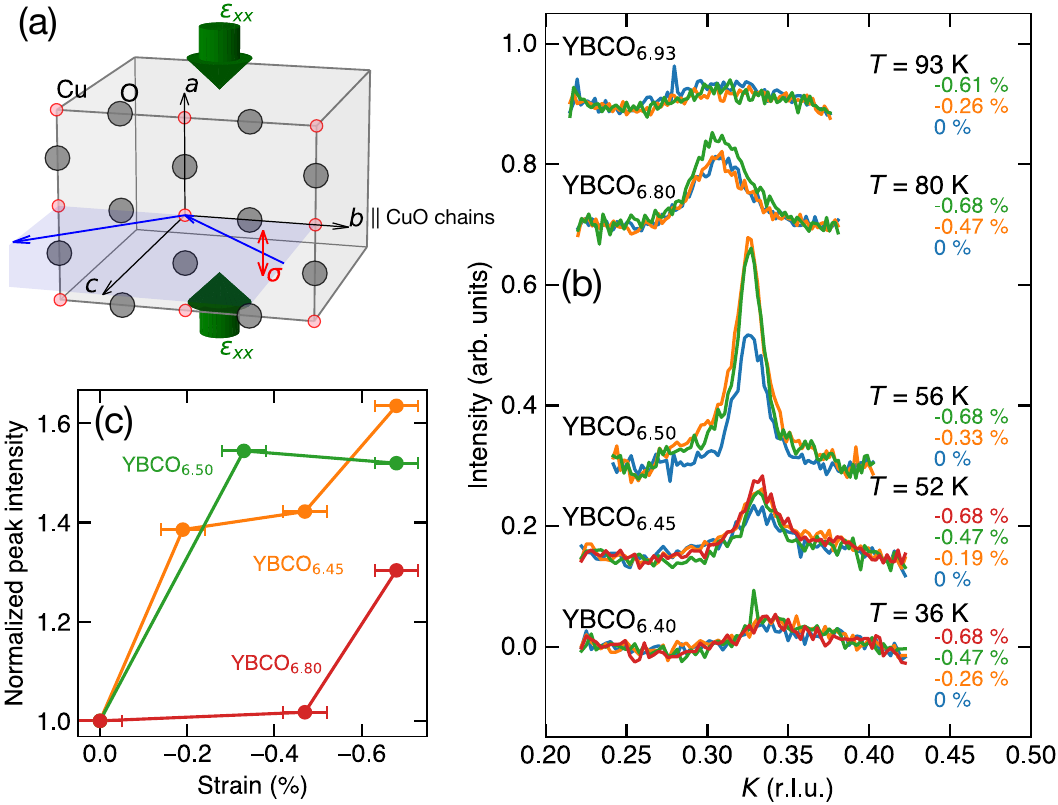}\\
  \caption{(a) Schematic geometry of the REXS measurements under stress applied along the $a$-axis. The scattering plane is parallel to the $bc$ plane, so that the $H$-component of the scattering vector is zero. (b) Strain dependence of the 2D CDW measured at $T=T_\text{c}$. A quadratic background was subtracted from the raw data. The curves for each doping level were vertically shifted for clarity. (c) Uniaxial stress dependence of the 2D CDW intensity. The intensity was normalized to the value in the absence of stress.}
  \label{fig:2dcdw}
\end{figure}

Figure \ref{fig:2dcdw} (b) displays background-subtracted REXS scans around the propagation vector of the 2D CDW ($(H,K) \sim (0, 0.32)$) of different YBCO$_{6+x}$ single crystals under uniaxial compression along the $a$-axis. The scans were collected at the unstrained $T_\text{c}$ of each sample, where the CDW correlation is maximized in the absence of magnetic fields (the components $H$, $K$, and $L$ of the wavevector, $Q$, are given in reciprocal-lattice units). The intensity of the 2D CDW along the $b$-axis is enhanced by $a$-axis strain in YBCO$_{6+x}$ ($x$ = 0.45, 0.50, and 0.80) (Fig. \ref{fig:2dcdw} (b,c)), akin to previous observations in YBCO$_{6.67}$ \cite{Kim1040,PhysRevLett.126.037002,Vinograd2024Using-stra} . However, we did not observe any discernible strain effect on the 2D CDW in the optimally doped sample YBCO$_{6.93}$ and very underdoped sample YBCO$_{6.40}$, where the charge order is very weak in the absence of strain. We emphasize that uniaxial stress was correctly applied and the expected strain level was reached in all samples investigated, including YBCO$_{6.40}$ and YBCO$_{6.93}$, as documented by the shift of the (002) Bragg peak, indicating the lattice deformation. Therefore the lack of strain effects in REXS scans at these specific doping levels is not due to technical issues affecting the efficiency of the uniaxial stress application.

\begin{figure}
  \includegraphics[width=\columnwidth]{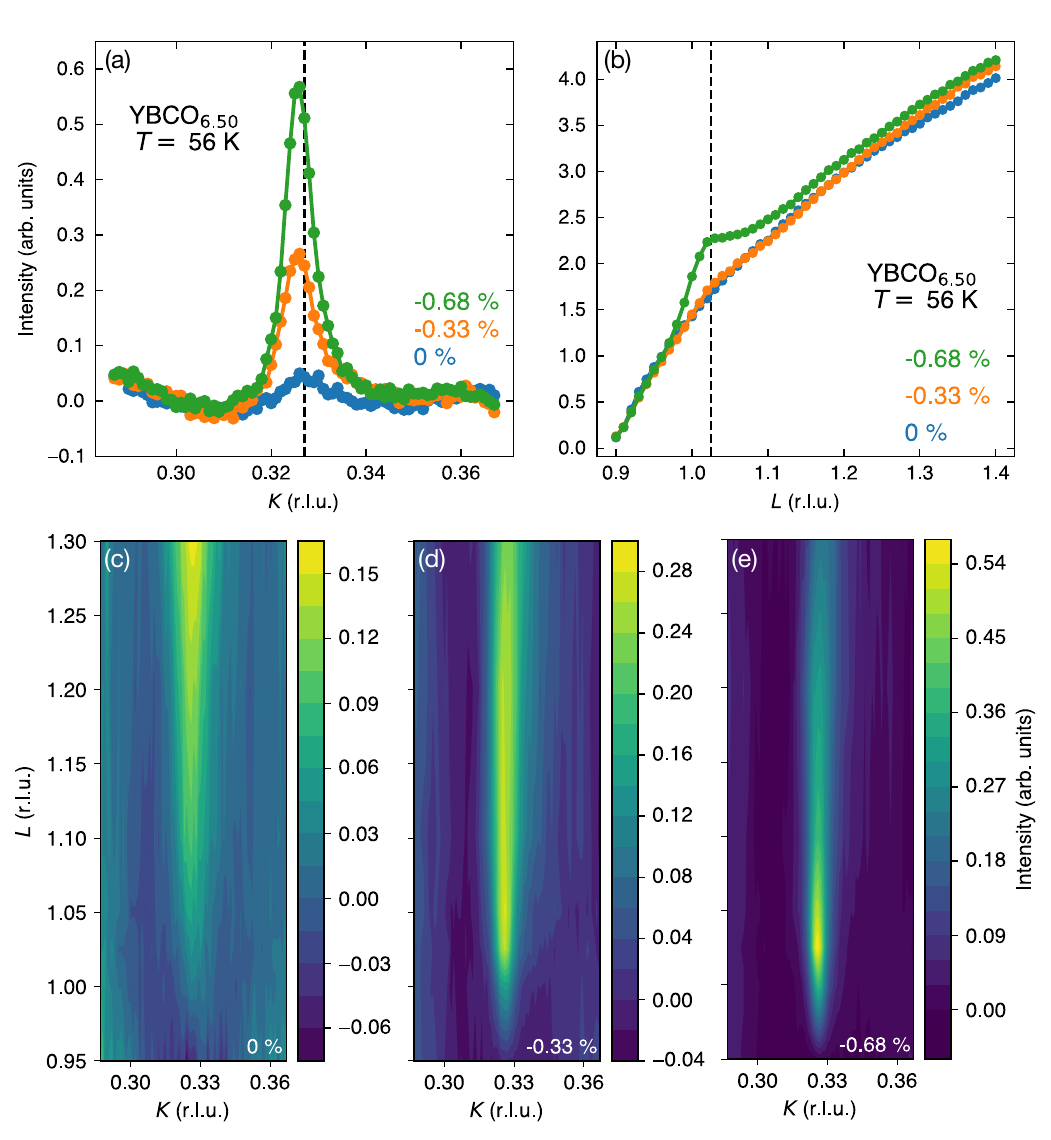}\\
  \caption{(a) Background-subtracted REXS intensity of YBCO$_{6.50}$ in the $K$ direction at ($H, L$) = (0, 1.025) under uniaxial stress measured at $T_\text{c}$ = 56 K. The dashed line indicates the $K$ value of panel b ($K$ = 0.327). (b) REXS intensity at ($H, K$) = (0, 0.327) along the $L$ axis under stress. The dashed line indicates the $L$ value of panel a ($L$ = 1.025). (c-e) Background-subtracted REXS intensity map in the $K$-$L$ plane in the absence of strain (c) and in the presence of strain of -0.33 \% (d) and -0.68 \% (e). }
  \label{fig:3dcdw_strain}
\end{figure}

Our central observation is summarized in the reciprocal-space maps of YBCO$_{6.50}$ shown in Fig. \ref{fig:3dcdw_strain} (c-e).
In the absence of strain, the quasi-2D nature of the CDW without significant correlations along the crystallographic $c$ direction is evident as a rod extending along the $L$ direction in the $K$-$L$ map (Fig. \ref{fig:3dcdw_strain} (c)). In the absence of strain, the intensity along the rod increases as $L$ approaches the half-integer value 1.5, indicating only very weak antiphase correlations between neighboring bilayers, as already reported in a number of previous studies \cite{Chang2016Magnetic-f,Kim1040}. At the intermediate strain level of -0.33 \%, the intensity distribution in the rod along the $L$ direction is different from the unstrained $K$-$L$ maps, as the intensity near the integer point $L$ = 1 becomes more prominent (Fig. \ref{fig:3dcdw_strain} \cor{(a) and} (d)). Figure \ref{fig:3dcdw_strain} (e) shows the data at the highest strain level (-0.68 \%) where a new peak-like feature emerges at ($K, L$) = (0.327, 1.025), signaling the 3D correlation of the CDW. 
This peak is not as intense as the one measured in the nonresonant hard x-ray scattering (NRXS) measurements where the diffraction signal of the 3D CDW was observed at ($K, L$) $\simeq$ (0.3, 7) \cite{Kim1040}. The disparity between NRXS and REXS is due to the different structure factors at the accessible momentum space positions, which is confined in REXS experiments by the limited photon energy at the Cu $L_3$ edge. As shown in Fig. \ref{fig:3dcdw_strain} (b), the intensity scan along the $L$ direction is strongly distorted, and the peak position ($L \sim 1.025$) is slightly greater than the integer value $L = 1$ where one expects to observe the 3D CDW Bragg peak. This is due to self-absorption effects caused by the grazing conditions for the outgoing photons, as discussed in the Supplemental Material of Ref. \onlinecite{PhysRevLett.126.037002}. The REXS peak in Fig. \ref{fig:3dcdw_strain} (b) is nearly identical to the 3D CDW signal from the elastic component of the RIXS data collected in the same scattering geometry \cite{PhysRevLett.126.037002}.

\begin{figure}
  \includegraphics[width=\columnwidth]{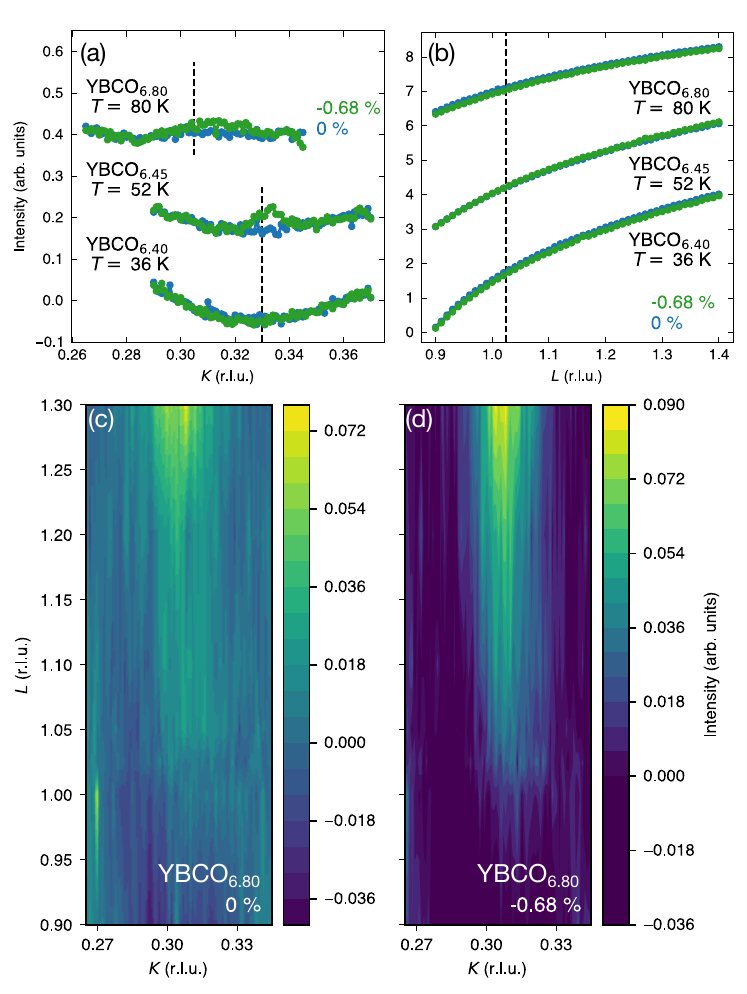}\\
  \caption{(a) Background-subtracted REXS intensity of YBCO$_{6.80}$, YBCO$_{6.45}$, and YBCO$_{6.40}$ in the $K$ direction at ($H, L$) = (0, 1.025) under strain measured at their $T_\text{c}$ (80 K, 52 K, and 36 K, respectively). The curves are vertically shifted for clarity. The range of the vertical axis is identical to that of Fig. \ref{fig:3dcdw_strain} (a) for comparison. The dashed line indicates the $K$ value of panel b ($K$ = 0.305 for YBCO$_{6.80}$ and 0.330 for YBCO$_{6.45}$ and YBCO$_{6.40}$). (b) REXS intensity at ($H, K$) = (0, 0.327) along the $L$ axis under strain. The curves are vertically shifted for clarity. The dashed line indicates the $L$ value of panel a ($L$ = 1.025). Note that the strained and unstrained curves of each doping level are almost identical. (c,d) Background-subtracted REXS intensity map of YBCO$_{6.80}$ in the $K$-$L$ plane in the absence of strain (c) and in the presence of strain of -0.68 \% (d). }
  \label{fig:3dcdw_strain2}
\end{figure}

In YBCO$_{6.80}$, the quasi-2D nature of the CDW is also evident in the $K$-$L$ map (Fig. \ref{fig:3dcdw_strain2} (c)). The rod extending along the $L$ direction becomes more intense under stress (Fig. \ref{fig:3dcdw_strain2} (d)) as already shown in Fig. 2 (b). However, no clear indication of the peak near $L = 1$ was observed even at the highest strain level ($\sim$ 0.68 \%), at which the 3D CDW was observed in YBCO$_{6.50}$ and YBCO$_{6.67}$. Analogous scans in YBCO$_{6.40}$ and YBCO$_{6.45}$ also show no clear peaks in the vicinity of the ordering vector of the 3D CDW (Fig. \ref{fig:3dcdw_strain2} (a,b)). Note that the REXS intensity at $L$ = 1.025 is enhanced by strain in YBCO$_{6.45}$ and possibly in YBCO$_{6.80}$ in Fig. \ref{fig:3dcdw_strain2} (a), but this is caused by a 'tail' of the 2D CDW peak centered at the half-integer $L$ as a consequence of the strain-enhanced 2D CDW (Fig. \ref{fig:2dcdw}) extending along the $L$ direction. In fact, no indication of the 3D CDW is noticeable in the $L$ scans (Fig. \ref{fig:3dcdw_strain2} (b)) unlike YBCO$_{6.50}$ (Fig. \ref{fig:3dcdw_strain} (b)).

\begin{figure}
  \includegraphics[width=\columnwidth]{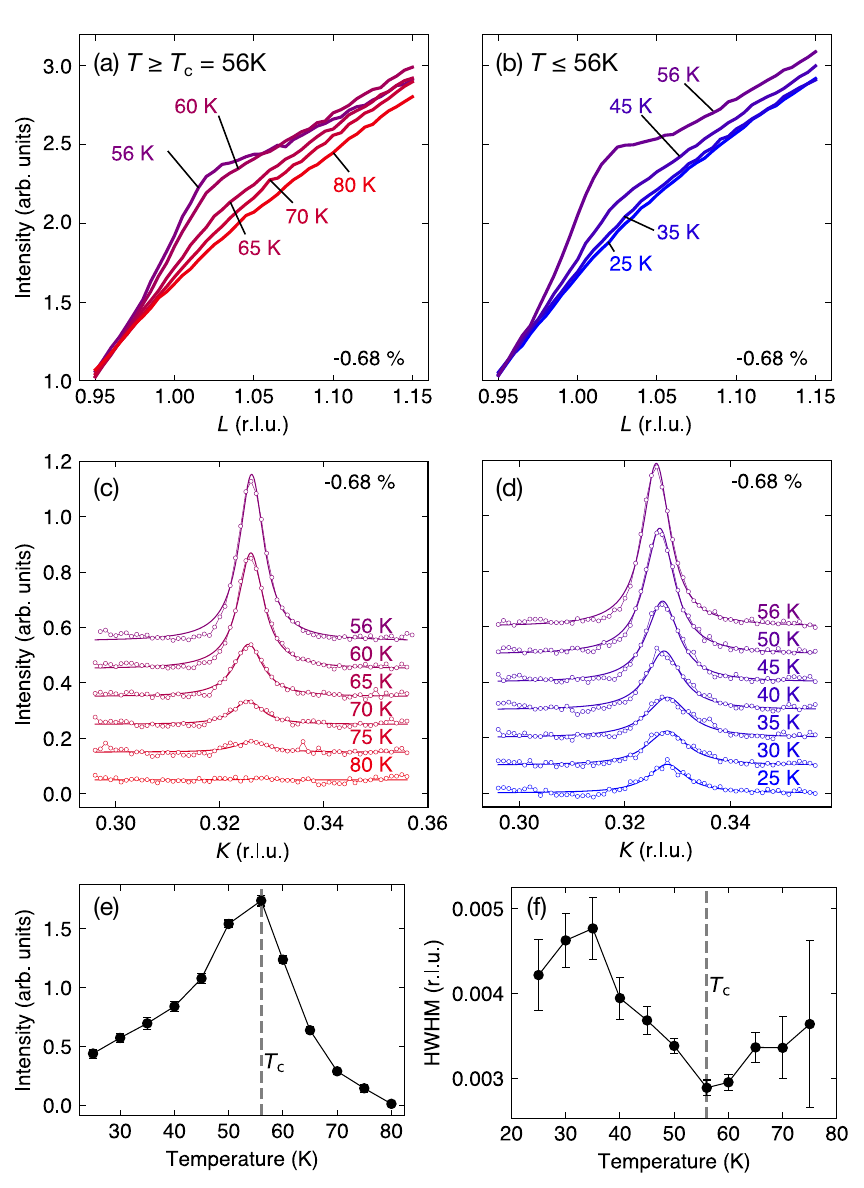}\\
  \caption{Temperature dependence of the strain-induced 3D CDW in YBCO$_{6.50}$. (a,b) Temperature dependence of the REXS intensity of $L$ scans collected around the same wave vectors as those in Fig. \ref{fig:3dcdw_strain} (b) above $T_\text{c}$ (a) and below $T_\text{c}$ (b). (c,d) Temperature dependence of the REXS intensity of $K$ scans collected around the same wave vectors as those in Fig. \ref{fig:3dcdw_strain} (a). The curves are vertically shifted for clarity. \cor{Both experimental data (open symbols) and fitted curves (solid curves) are shown.} (e,f) Results of fits of the $K$ scans shown in the panels (c,d) to Lorentzian profiles. The integrated intensity (e) and half-width-at-half-maximum (HWHM) (f) are shown. Error bars are standard deviations of the fits; the error bars in (e) are smaller than the symbol size. The vertical dashed line indicates the unstrained $T_\text{c}$.}
  \label{fig:3dcdw_tdep}
\end{figure}

To investigate the relationship between the CDW and superconductivity in YBCO$_{6.50}$, we measured the temperature dependence of the strain-induced 3D CDW with fixed strain (-0.68 \%). Figure \ref{fig:3dcdw_tdep} (a,b) ((c,d)) shows the temperature dependence of the REXS scans along the $L$ ($K$) direction around $Q =$ (0, 0.327, 1.025). The intensity is maximized at the unstrained $T_\text{c}$, which implies a strong competition with superconductivity (note that the strain dependence of $T_\text{c}$ in YBCO$_{6.50}$ has not been measured, {but in YBCO$_{6.67}$ the decrease in $T_\text{c}$ by the onset strain of the 3D CDW is only $\sim$5 K \cite{PhysRevB.106.184516}}.). At this doping and strain level, the onset temperature of the 3D CDW is $\sim$80 K (Fig.  \ref{fig:3dcdw_tdep} (e)), whereas the one of the 2D CDW  is $\sim$125 K \cite{PhysRevB.90.054513}.  The sharpness of the REXS peaks of the 3D CDW along $K$ enables us to fit the data by a single Lorentzian. The temperature dependence of the fitting parameters are displayed in Fig. \ref{fig:3dcdw_tdep} (e, f). The competition between the 3D CDW and superconductivity is clear from the fitting results. Nonetheless, a non-negligible 3D CDW with a quarter of the intensity at $T_\text{c}$ persists at the lowest temperature. On the other hand, half of the maximum intensity of the 2D CDW at $T_\text{c}$ in YBCO$_{6.51}$ previously studied in the same setup persists at the lowest temperature \cite{PhysRevB.90.054513}. Therefore the strain-induced 3D CDW appears more strongly suppressed with superconductivity than the 2D CDW as in the case of YBCO$_{6.67}$ \cite{Kim1040}.

\section{\label{sec:level4}Discussion}

We first discuss the strain-induced 3D CDW, which was observed only in YBCO$_{6.50}$ in the present study. All major characteristics of this state are closely similar to the one in YBCO$_{6.67}$ reported in previous studies, despite the different oxygen ordering patterns in the CuO chain layers (Ortho-II and Ortho-VIII in YBCO$_{6.50}$ and YBCO$_{6.67}$, respectively (Table \ref{table:YBCO_REXS})). The 3D CDW was not observed in YBCO$_{6.40}$ and YBCO$_{6.45}$, both of which exhibit the same Ortho-II order as YBCO$_{6.50}$. The oxygen-order insensitive stress response confirms that details of the ordering pattern of the CuO chains are not crucial for the formation of the 3D CDW. Consistent with this observation, the 3D CDW intensity exhibits a resonant enhancement only at the absorption energy of planar Cu atoms in the CuO$_{2}$ planes in YBCO$_{6.67}$ \cite{PhysRevLett.126.037002}. As the 2D CDW of YBCO$_{6.67}$ also barely depends on the chain order \cite{PhysRevLett.113.107002}, we conclude that the structure of the CuO chains plays at most a minor role for the formation of both forms of charge order. Note that this is at odds with the 3D charge order observed in YBCO$_{6+x}$ thin films, where the photon energy dependence of the diffraction signal shows a contribution from the Cu atoms in the CuO chain layers and the weak temperature dependence suggests structural contributions not primarily driven by instabilities of the valence-electron system  \cite{Bluschke2018Stabilizat}.

Whereas there are only minor differences between the temperature dependencies of the strain-induced 3D CDWs at both doping levels, the temperature range of the strain-induced 3D CDW is clearly distinct from the one of the magnetic-field-induced 3D CDW, which was observed only below a temperature of 50 K (which is comparable to the superconducting $T_\text{c}$) in the entire doping regime \cite{Gerber949,Chang2016Magnetic-f}. By contrast, our data show an onset of the strain-induced 3D CDW well above $T_\text{c}$.

The strain-induced 3D CDW was not observed in YBCO$_{6.45}$ and YBCO$_{6.80}$, despite the enhancement of the 2D CDW. So far the magnetic-field-induced 3D CDW was also reported only in a limited doping regime: between YBCO$_{6.48}$ and YBCO$_{6.67}$ in x-ray scattering \cite{Gerber949,Jang14645,Chang2016Magnetic-f} and between YBCO$_{6.56}$ and YBCO$_{6.68}$ in nuclear magnetic resonance (NMR) experiments \cite{Wu2015Incipient-}. Ultrasound measurements indicated a thermodynamic phase transition into the magnetic-field-induced 3D CDW at higher doping levels including YBCO$_{6.75}$ and YBCO$_{6.79}$, which however require larger magnetic fields of 25 and 35 T, respectively \cite{Laliberte2018High-field}. Therefore, a plausible reason for the absence of the strain-induced 3D CDW in YBCO$_{6.45}$ and YBCO$_{6.80}$ is that the maximum stress we achieved was smaller than the onset stress of the 3D CDW at these doping levels. Much higher uniaxial pressure levels of -3.2 GPa (which corresponds to a compressive strain of $\sim$ -2 \% for YBCO$_{6+x}$) were recently achieved in experiments on Sr$_2$RuO$_4$. This required a more elaborate sample shaping method, which could possibly be extended to the cuprates in future studies \cite{Jerzembeck2022The-superc}.

The three-dimensionally long-range ordered, uniaxial CDW should result in pronounced transport anisotropies, which will be interesting to explore, in analogy to recent work on strain-induced anisotropies of the resistivity and Hall effect associated with the 2D CDW \cite{Nakata2022Normal-sta}. On general grounds, one expects these anisotropies to persist in a fluctuation regime at temperatures exceeding the transition temperature for 3D CDW order. At zero strain, transport \cite{PhysRevB.92.224502}, magnetization \cite{Sato2017Thermodyna}, and neutron scattering \cite{Hinkov597,1367-2630-12-10-105006} experiments have been interpreted as evidence of a separate transition into a ``nematic" phase where the $C_4$ rotational symmetry is broken, but translational symmetry is preserved. Note that the orthorhombic crystal structure of YBCO$_{6+x}$ implies that the electronic nematic transition is necessarily transformed into a crossover, but the notion of electronic nematicity may remain useful if electronic mechanisms dominate the macroscopic response, analogous to the iron-based superconductors \cite{Fernandes2014What-drive}. The extent to which conduction electrons can affect elastic properties was also recently demonstrated in Young's modulus experiments on Sr$_2$RuO$_4$ \cite{Noad2023Giant-latt}.

In this context, we note that sharp Kohn anomalies of high-energy Cu-O-Cu bond bending and stretching phonons have been observed along the $b$-axis (but not the $a$-axis) in underdoped and in optimally doped YBCO$_{6+x}$ \cite{PhysRevLett.89.037001,PhysRevLett.107.177004}. Since the in-plane anisotropy of these anomalies matches the one of the 3D CDW, they might be regarded as precursors of this phase. Interestingly, a Kohn anomaly with a closely similar wavevector was recently also observed in lanthanum-based superconductors, suggesting that the CDW instability may be generic to the cuprates \cite{PhysRevX.8.011008}. Note that less anisotropic anomalies of low-energy phonons have been discussed in conjunction with the quasi-2D CDW \cite{Le-Tacon2014Inelastic-,doi:10.7566/JPSJ.90.111006}.  The behavior of the Kohn anomalies and the ``electronic nematic" response at high strain levels is an interesting subject of future research.

We now turn to the stress response of the 2D CDW, which (unlike the 3D CDW) is observed both along the $a$- and along the $b$-axis. A comprehensive REXS investigation of YBCO$_{6.67}$ demonstrated that this phenomenology is due to the coexistence of two uniaxial CDW domains and revealed a symmetric stress response: pressure along $a$ enhances the CDW along $b$, and vice versa \cite{PhysRevLett.126.037002}. We have therefore confined our doping-dependent study to the former geometry. In YBCO$_{6.50}$, the sample with the most intense, longest-range 2D-CDW correlations, the diffraction peak is strongly enhanced even at low stress levels (Fig. 2b), mirroring the observations in YBCO$_{6.67}$ \cite{PhysRevLett.126.037002}. The pressure scale is lower than the one required to induce the 3D CDW, because the effect is due to domain selection rather than a pressure-induced phase transition. Similarly, uniaxial spin and charge order in lanthanum-based cuprates was also found to be susceptible to strain of the order of 0.01 - 0.1 \%, much smaller than the strain inducing the 3D CDW in YBCO$_{6+x}$ \cite{PhysRevLett.128.207002,Simutis2022Single-dom,Wang2022Uniaxial-p,PhysRevResearch.3.L022004}. 

A stress-induced enhancement of the 2D-CDW diffraction peak intensity is still clearly observable in the YBCO$_{6.45}$ and YBCO$_{6.80}$ samples, where the CDW correlations are weaker than around $p = 1/8$, but no longer discernible in YBCO$_{6.40}$ and YBCO$_{6.93}$, which are close to the borders of the 2D CDW stability range (Fig. 2b). The doping-dependent stress response is thus symmetric around $p = 1/8$, as expected if the 2D CDW is modified directly by strain rather than by secondary effects such as a strain-induced modification of the hole density. The large momentum widths of the CDW peaks in YBCO$_{6.40}$ and YBCO$_{6.93}$ indicate a smaller correlation length comparable to those observed in more disordered cuprates such as  Bi$_2$Sr$_2$CaCu$_2$O$_{8+\delta}$ \cite{Frano_2020}. This finding suggests that the CDW correlations in these samples may be pinned to structural defects and hence less susceptible to external control parameters. In agreement with this scenario, the CDW diffraction features in YBCO$_{6+x}$ samples near the stability range of the 2D CDW also exhibit less pronounced anomalies at the superconducting $T_\text{c}$ \cite{PhysRevB.90.054513}.

\begin{figure}
  \includegraphics[width=\columnwidth]{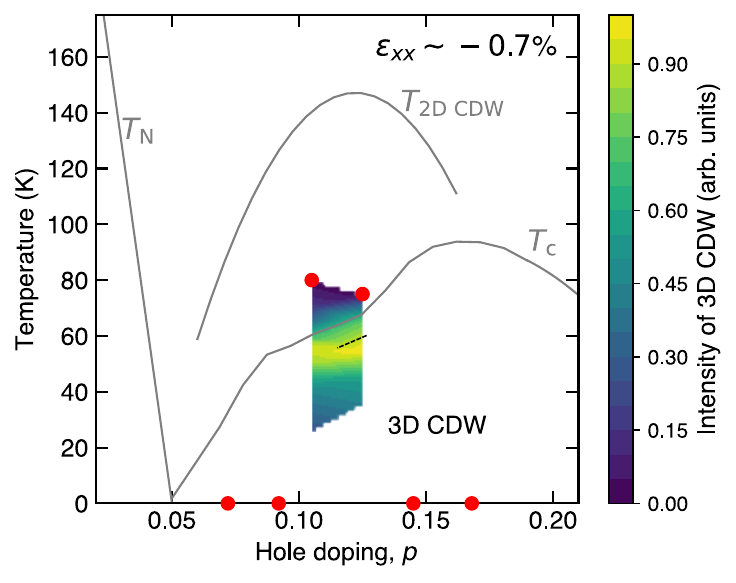}\\
  \caption{Phase diagram of YBCO$_{6+x}$ under uniaxial compression along the $a$-axis. The lines indicate the N\'eel temperature, superconducting $T_\text{c}$, and onset temperature of the 2D CDW at ambient pressure. The dashed line marks $T_\text{c}$ of YBCO$_{6.67}$ \cite{PhysRevB.106.184516}, and the red points indicate the onset temperatures of the 3D CDW  under ~0.7\% compression. The color map shows the integrated intensity of the 3D CDW of YBCO$_{6.50}$ and YBCO$_{6.67}$ investigated in the present work and reported in Ref. \onlinecite{PhysRevLett.126.037002}, respectively. The intensity was normalized to its maximum value. }
  \label{fig:5_phase_diagram}
\end{figure}

\section{\label{sec:level5}Conclusion and outlook}

In conclusion, we have investigated the nature of 2D and 3D CDW correlations in the CuO$_2$ planes under $a$-axis compression up to $\sim$ 0.7 \% by means of REXS measurements over a large range of doping levels. The amplitude of the stress response of both forms of CDW order is maximal around $p \sim 1/8$ (Fig. \ref{fig:5_phase_diagram}), thereby apparently mirroring the doping-dependent amplitude, correlation length, and onset temperature of the CDW at zero stress \cite{PhysRevB.90.054513}. 

The 2D CDW correlations are characterized by coexisting finite-sized domains propagating along the $a$- and $b$-axes. External uniaxial pressure primarily modulates the volume fractions of both domains. By contrast, the long-range ordered 3D CDW is generated via a stress-induced phase transition and only observed along $b$, at least at the stress levels accessible in our study. The in-plane anisotropy of this phenomenon likely reflects the influence of the electronic anisotropy of YBCO$_{6+x}$ induced by hybridization of chain and plane bands \cite{ANDERSEN199031}, which is more profound than the expected influence of the minor difference between the lengths of the $a$- and $b$-axes. The temperature dependence of the 3D CDW amplitude indicates a strong competition with superconductivity. \cor{These results should be viewed in the context of experiments on other superconducting quantum materials under the application of uniaxial stress. In particular} we refer to the stress-induced spin-density wave transition in superconducting Sr$_2$RuO$_4$, which has been attributed to stress tuning of one of its Fermi surfaces across a van-Hove singularity \cite{Grinenko2021Split-supe,Li2022Elastocalo}. Interestingly, the converse effect was recently reported for La$_{2-x}$Ba$_x$CuO$_4$ with $x = 0.115$, where uniaxial pressure obliterates long-range spin and charge order and restores the uniform superconducting state \cite{PhysRevLett.125.097005}.

Our findings call for a variety of further experiments on YBCO$_{6+x}$ (as well as La-, Hg- or Bi-based cuprates with more closely tetragonal crystal structures) under high in-plane uniaxial stress, including experiments under tensile and higher compressive stress for a more complete exploration of the phase diagram, as well as complementary measurements of the Fermi surface and band dispersions (for instance via quantum oscillations or angle-resolved photoemission) under the same conditions. Following up on recent experiments at lower strain levels \cite{Nakata2022Normal-sta}, experiments combining REXS with {\it in-situ} transport and thermodynamic measurements have the potential to yield new insight into the influence of atomic-scale charge correlations on the macroscopic electronic properties of YBCO$_{6+x}$ and other cuprates.

\section*{Acknowledgements}
\cor{We acknowledge Michele Tortora and Gaston Garbarino for support during strain calibration tests at ID27 at the ESRF.} Self-flux growth was performed by the Scientific Facility ‘Crystal Growth’ at Max Planck Institute for Solid State Research, Stuttgart, Germany. S.M.S. acknowledges funding by the DFG – Projektnummer 44123158. M.LT., T.L., S.M.S., and C.W.H. acknowledge the funding of the Deutsche Forschungsgemeinschaft (DFG, German Research Foundation), projects 422213477 (TRR 288 projects B03). C.W.H. acknowledges support from the Engineering and Physical Sciences Research Council (U.K.) (EP/X01245X/1).

\section*{\label{sec:levelAPP}Appendix A: Strain scale}

\begin{figure}[H]
    \begin{center}
    \includegraphics[width=\columnwidth]{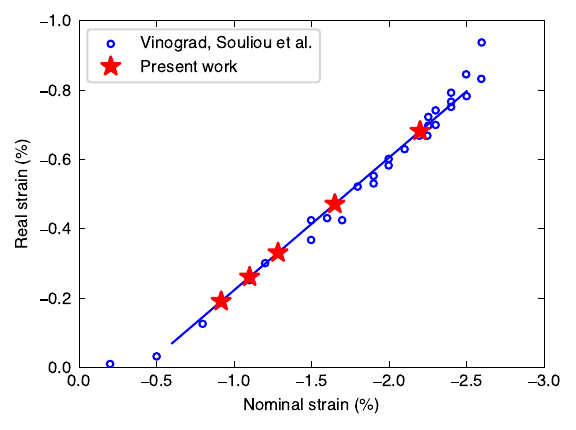}
    \caption{Real strain measured by x-ray diffraction \cite{Vinograd2024Using-stra} versus nominal strain inferred from a capacitive sensor reading in YBCO$_{6+x}$. The blue data points are reproduced from Ref. \onlinecite{Vinograd2024Using-stra}. The blue line is a linear fit to these data. The red stars indicate the data points in the present study.}
    \label{fig:strain}
    \end{center}
\end{figure}

Here we explain the calibration of the strain value in the present REXS study. In prior work on YBCO$_{6+x}$, the strain level was read off from a capacitive sensor of the sample length in the strain device. However, this reading (hereafter referred to as ``nominal strain") has turned out to be inaccurate because of incomplete strain transmission from the device to the sample. Although an attempt was made to model the magnitude of the strain, it was overestimated. In a recent study, Vinograd, Souliou {\it et al.} directly determined the actual strain level in the sample (hereafter referred as ``real strain") from measurements of the lattice parameters by high-resolution hard x-ray diffraction in a device nearly identical to ours, and on needle-shaped samples with the same aspect ratio as those in the present work \cite{Vinograd2024Using-stra}. We used a linear fit to the data of Ref. \onlinecite{Vinograd2024Using-stra} (Fig. \ref{fig:strain}) to convert the capacitive sensor readings of our device to the actual strain. Despite  some systematic error associated with strain transmission by the glue used to hold the samples in place (which may differ from sample to sample), we note that \cor{using the calibration described above, the experimental results reported here are compatible with those reported in Ref. \onlinecite{Vinograd2024Using-stra}, i.e., they show the same evolution of 2D and 3D CDW vs. strain. Specifically, the 2D CDW signal displays the same intensity enhancement upon uniaxial pressure application and the 3D CDW, at the doping levels where it appears, displays the same strain threshold in both resonant and nonresonant scattering experiments, thus demonstrating the internal consistency between datasets. Within each strain run, we have always been able to distinguish and reproduce strain steps as small as $\sim$0.1 \% compression on the real strain scale. Fig. \ref{fig:2dcdw}(c) therefore shows a conservative error bar of $\pm $ 0.05\%.}

\begin{figure}[H]
    \begin{center}
    \includegraphics[width=0.9\columnwidth]{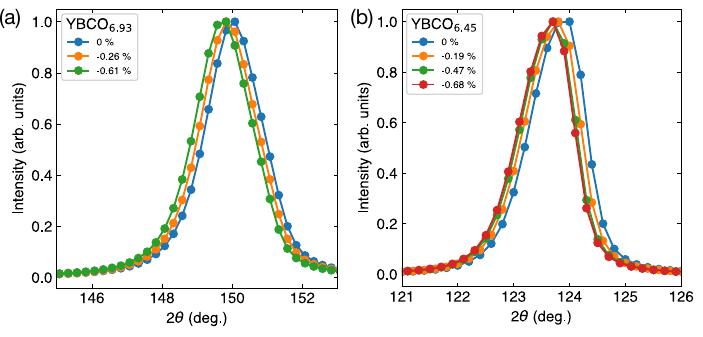}
    \caption{(002) Bragg peak shifts of YBCO$_{6+x}$ in response to strain. The data were collected at the respective $T_\text{c}$. (a) YBCO$_{6.93}$. (b) YBCO$_{6.45}$. Note that the photon energy of the x-ray was set to 930 eV and 1200 eV for the panels (a) and (b), respectively.}
    \label{fig:Bragg}
    \end{center}
\end{figure}

As a secondary diagnostic, we have monitored the lattice Bragg reflection peak (002) during the REXS measurements and confirmed that this peak shifts monotonically with in-plane uniaxial stress (Fig. \ref{fig:Bragg}). However, it is difficult to quantitatively estimate the in-plane strain from out-of-plane Bragg reflection peaks, which are only affected by strain via Poisson's ratio. Moreover, (002) is a low-order reflection with a width that is large ($\sim 5 ^\circ$) in soft x-ray diffraction compared to the peak shift under strain levels of order $<-1\%$ ($\sim 0.1 ^\circ$). 

\section*{\label{sec:APP2}Appendix B: Raw data of 2D CDW}

\begin{figure}[t]
    \begin{center}
    \includegraphics[width=0.8\columnwidth]{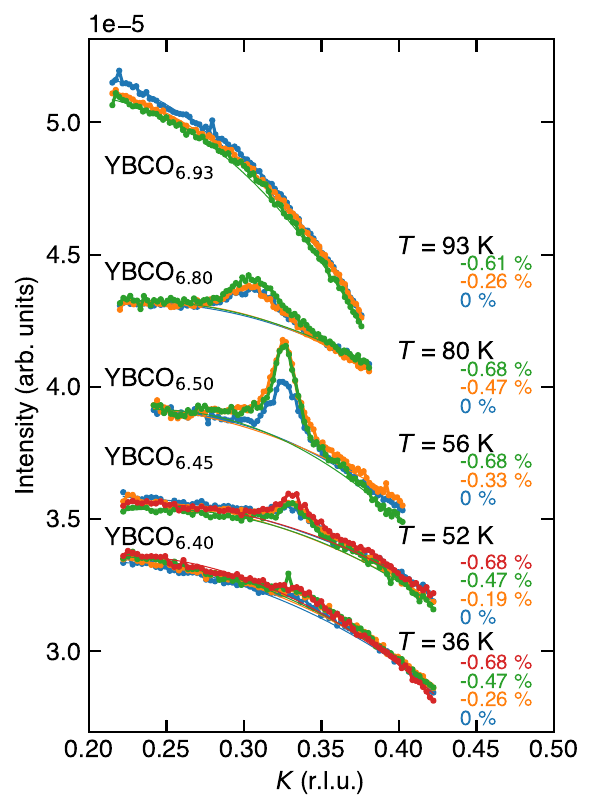}
    \caption{Raw data of Fig. 2(b).}
    \label{fig:rawdata}
    \end{center}
\end{figure}

\cor{Figure \ref{fig:rawdata} displays raw data of Fig. \ref{fig:2dcdw} (b) before subtracting the background. As can be seen in the raw data, the background is not linear along $K$. We therefore used fits to quadratic polynomials, following common practice (see e.g. Ref. \onlinecite{PhysRevB.90.054513}). The same background subtraction was made for all the REXS data along $K$ presented here.}

\section*{\label{sec:APP3}Appendix C: Strain homogeneity}

\begin{figure}[H]
    \begin{center}
    \includegraphics[width=0.9\columnwidth]{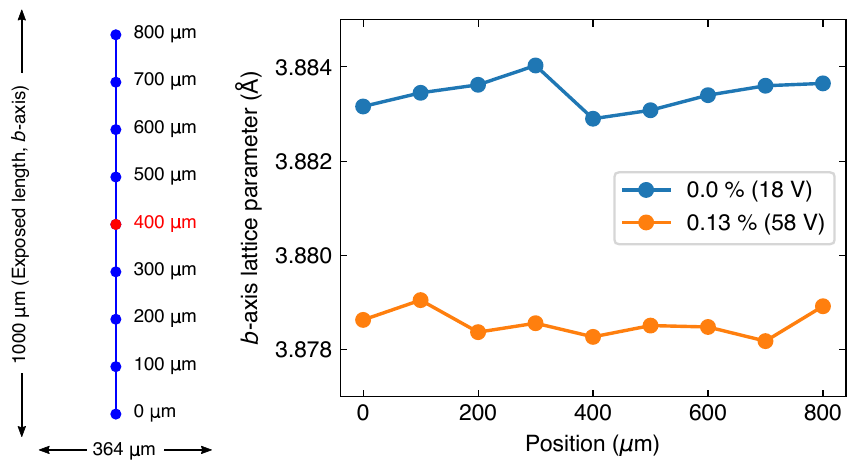}
    \caption{\cor{Lattice parameter $b$ of YBCO$_{6+x}$ as a function of x-ray spot position in the presence and absence of strain. The lattice parameter at every position of the sample was inferred from the Bragg reflection (034) collected at various positions of the sample along the uniaxial stress direction. The x-ray diffraction data were taken at 70 K.}}
    \label{fig:ESRF}
    \end{center}
\end{figure}

\cor{Prior to this study we did extensive tests to map the strain over several points across YBCO needles at the high resolution diffraction beamline ID27 of the European Synchrotron ESRF. The tests, as shown in Fig. \ref{fig:ESRF}, demonstrate that strain is very homogenous, consistently with the results of Ref. \onlinecite{doi:10.1063/1.4881611}. Combining the information from simulations and tests, it is evident that focusing on the central region of the needle allows one to be extremely safe in terms of strain homogeneity.}

%

\end{document}